\documentclass[preprint,12pt]{elsarticle}

\usepackage{amssymb}
\usepackage{epsfig}
\usepackage{graphics}
\usepackage{subfigure}
\usepackage{graphicx}
\usepackage{rotating}

\journal{Physics Letters B}

\begin{document}

\begin{frontmatter}



\title{Search for $\nu_\mu\rightarrow\nu_\tau$ oscillation with the OPERA experiment in the CNGS beam}

\author[label31]{N.~Agafonova}
\author[label19]{A.~Aleksandrov\fnref{onleave}}
\author[label38]{O.~Altinok}
\author[label34]{A.~Anokhina}
\author[label26]{S.~Aoki}
\author[label35]{A.~Ariga}
\author[label35]{T.~Ariga}
\author[label5]{D.~Autiero}
\author[label36]{A.~Badertscher}
\author[label33]{A.~Bagulya}
% ...

\fntext[onleave]{On leave of absence from LPI-Lebedev Physical Institute of the Russian Academy of Sciences, 119991 Moscow, Russia}
\fntext[giorgini]{Now at INAF/IASF, Sezione di Milano, I-20133 Milano, Italy}
\fntext[gustavino]{Now at Dipartimento di Fisica dellÕUniversitˆ di Roma ÒLa SapienzaÓ and INFN, I-00185 Roma, Italy}
\fntext[korea]{Now at Chonnam National University, Korea}
\fntext[mauri]{Now at INFN - Laboratori Nazionali di Frascati dellÕINFN, I-00044 Frascati (Roma), Italy}
\fntext[park]{Now at Asan Medical Center, 388-1 Pungnap-2 Dong, Songpa-Gu, Seoul 138-736, Korea}
\fntext[scotto]{Now at SUBATECH, CNRS/IN2P3, F-44307 Nantes, France}

\cortext[ca]{Corresponding Author}
\address[label31]{INR-Institute for Nuclear Research of the Russian Academy of Sciences, RUS-327312 Moscow,  Russia}
\address[label19]{INFN Sezione di Napoli, I-80125 Napoli, Italy}
\address[label38]{METU-Middle East Technical University, TR-06532 Ankara, Turkey}
\address[label34]{(MSU SINP) Lomonosov Moscow State University Skobeltsyn Institute of Nuclear Physics , RUS-329992 Moscow, Russia}
\address[label26]{Kobe University, J-657-8501 Kobe, Japan}
\address[label35]{Albert Einstein Center for Fundamental Physics, Laboratory for High Energy Physics (LHEP), University of Bern, CH-3012 Bern, Switzerland}
\address[label5]{IPNL, Universit\'e Claude Bernard Lyon I, CNRS/IN2P3, F-69622 Villeurbanne, France}
\address[label36]{ETH Zurich, Institute for Particle Physics, CH-8093 Zurich, Switzerland}
\address[label33]{LPI-Lebedev Physical Institute of the Russian Academy of Science, RUS-119991 Moscow, Russia}
% ...

\begin{abstract}
The OPERA neutrino experiment in the underground Gran Sasso Laboratory (LNGS) was designed to perform the first detection of neutrino oscillations in direct appearance mode in the $\nu_\mu\rightarrow\nu_\tau$ channel, the  $\nu_\tau$ signature being the identification of the  $\tau$-lepton created in its charged current interaction. The hybrid apparatus consists of a large mass emulsion film/lead target complemented by electronic detectors. It is placed in the high energy long-baseline CERN to LNGS neutrino beam (CNGS) 730 km away from the neutrino source. The observation of a first  $\nu_\tau$ candidate event was reported in 2010. In this paper, we present the status of the experiment based on the analysis of the data taken during the first two years of operation (2008-2009). The statistical significance of the one event observed so far is then assessed.

\end{abstract}

\begin{keyword}


\end{keyword}

\end{frontmatter}


\section{Introduction}
\label{intro}
Neutrino oscillations were first predicted nearly 50 years ago \cite{pontecorvo}  and definitely established in 1998 by the Super-Kamiokande experiment with atmospheric neutrinos \cite{SK}. Several other experiments carried out in the last decades with atmospheric, solar, reactor and accelerator neutrinos have established our current understanding of neutrino mixing and oscillations (see e.g. \cite{review}  for a review). In particular, the depletion in the $\nu_\mu$ neutrino flux through oscillation observed by several atmospheric neutrino experiments \cite{SK,K} was confirmed by two accelerator experiments \cite{K2K}. The fact that $\nu_\mu\rightarrow\nu_e$ oscillation cannot be the dominant channel has been indirectly confirmed by two nuclear reactors experiments \cite{CHOOZ}. An indication of $\nu_e$ appearance in a $\nu_\mu$ beam with a statistical significance of 2.5 $\sigma$ has recently been published by the T2K experiment \cite{T2K}. However, a direct flavour transition has not yet been established where the oscillated neutrino is identified by the charged lepton created in its charged current (CC) interaction. The appearance of $\nu_\tau$ in an accelerator $\nu_\mu$ beam will unambiguously prove that $\nu_\mu\rightarrow\nu_\tau$ oscillation is the dominant transition channel for the neutrino atmospheric sector. This is the main goal of the long-baseline OPERA experiment \cite{operaold,proposal1}, with its detector exposed in the Gran Sasso Underground Laboratory (LNGS) to the high energy CERN CNGS neutrino beam \cite{CNGS}.
 
The detection of CNGS neutrino interactions in OPERA was reported in \cite{operafirst} and the observation of a first $\nu_\tau$ candidate event was presented in \cite{operacandidate}. In this paper we summarize the major improvements brought to the analysis chain and to the Monte Carlo simulations. They mainly concern the evaluation of the efficiencies and the reduction or better control of the physics backgrounds. The event statistics acquired during the first two full years of data taking in 2008 and 2009 is used for the studies reported here. The statistical significance of the observation of one $\nu_\tau$ candidate event reported in \cite{operacandidate} is re-assessed.

\section{The OPERA detector and the CNGS beam}
\label{chap2}
The challenge of the OPERA experiment is to achieve the very high spatial accuracy required for the detection of $\tau$ leptons (whose decay length is
of the order of 1 mm in this experiment) inside a large-mass active target. The hybrid detector \cite{operadetector} is composed of two identical Super Modules (SM), each consisting of an instrumented target section of a mass of about 625 tons followed by a magnetic muon spectrometer. A target section is a succession of walls filled with elements called ÒbricksÓ, interleaved with planes of scintillator strips composing the Target Tracker (TT) that triggers the read-out and allows localizing neutrino interactions within the target. A brick is an Emulsion Cloud Chamber (ECC) module consisting of 56 1-mm thick lead plates interleaved with 57 nuclear emulsion films. It weighs 8.3 kg and its thickness corresponds to 10 radiation lengths along the beam direction. Tightly packed removable doublets of emulsion films called Changeable Sheets (CS) are glued to the downstream face of each brick. They serve as interfaces between the TT planes and the bricks to facilitate the location of neutrino interactions. Large brick handling ancillary facilities are used to bring emulsion films from the target up to the automatic scanning microscopes in Europe and Japan. Extensive information on the OPERA detector and facilities is given in \cite{operadetector,ESS,facilities}. 

OPERA is exposed to the long-baseline CNGS $\nu_\mu$ beam \cite{CNGS}, 730 km away from the source. The beam is optimized for the observation of $\nu_\tau$CC interactions. The average neutrino energy is $\sim$17 GeV. In terms of interactions, the $\bar{\nu}_\mu$ contamination is 2.1\%, the $\nu_e$ and $\bar{\nu}_e$ contaminations are together lower than 1\%, while the number of prompt $\nu_\tau$  is negligible. 

\section{Location of neutrino interactions}
\label{selection}
The expected number of neutrino events registered in the target volume is 850 per $10^{19}$ protons on target (p.o.t.) per 1000 tons. A 10\% error is assigned to this number resulting from uncertainties on the neutrino flux and interaction cross-sections. During the 2008 and 2009 runs the average target mass was 1290 tons, of
which 8.6\% of dead material other than lead plates and emulsion films, for a total number of $1.78\times10^ {19}$  p.o.t in 2008 and $3.52\times 10^{19}$ p.o.t. in 2009.

With a trigger efficiency of 99\% the total numbers of triggers in 2008 and 2009 on-time with the beam amount to 10121 and 21455, respectively. About 85\% of these are due to particles entering the target after being emitted in neutrino interactions occurring in the rock surrounding or in material inside the LNGS cavern. This component, hereafter called Òexternal eventsÒ, has been disentangled during the 2008 run by performing a visual inspection of all on-time events. Events with topologies consistent with charged particles entering the detector or with low energy interaction of neutrons and $\gamma$-rays, mimicking a neutral current (NC) interaction, were discarded from the sample. For the 2008 data, stringent criteria were used guaranteeing a high level of purity at the cost of some inefficiency for very low activity events. A total of 1698 events was retained and constitute the 2008 sample. In 2009, events compatible with occurring in the target volume have been selected by an automatic algorithm \cite{opcarac,eledet} developed on the basis of the experience acquired with the 2008 sample. Automating the procedure reduces the human workload required by the visual inspection. It also aims at reaching higher efficiency for a specific category of signal events: $\tau$ emitted in Quasi-Elastic (QE) interactions and decaying to an electron. These have a topology very similar to that of NC-like external events. The drawback is an increased contamination by external events that must be measured.

A Monte Carlo simulation including both external events and interactions occurring inside the targets well reproduces the experimental data for what concerns the event rates and the position distribution of the events inside the targets, validating the automatic selection algorithm. This is in particular true for the low energy external NC-like events which tend to accumulate close to the borders of the targets. The efficiency of the algorithm is estimated by simulation to be 96.2\% for the CC events and 86.3\% for the NC events while the contamination by external events, lower than 1\% for CC events, is 23.3\% for NC events. By applying the automatic selection algorithm on the 2009 sample 3629 events were expected from the simulation and 3693 were eventually retained from the experimental data. The sample was further reduced to 3557 events after visual inspection of the event displays.

Data from the electronic detectors associated with the 5255 events reconstructed to have occurred inside the target volume were processed by a software algorithm that selects the brick with the highest probability to contain the neutrino interaction vertex. The brick so designated is removed from the target, the CS is detached and its films are searched for tracks compatible with the electronic data to verify the brick selection. In case this search is unsuccessful, the brick is equipped with a fresh CS and reinserted into the target. A second brick is then extracted according to its probability to contain the vertex complemented by a visual inspection of the event display. In case the search is successful the brick is dismounted and the emulsion films are developed and dispatched to the scanning laboratories. All tracks measured with high precision in the CS films are sought for in the most downstream films of the brick. These tracks are then followed back until they are not found in three consecutive films. A volume is then scanned around their stopping point in order to localize the interaction vertex.

\section{Search for decay topologies: observation of a first $\nu_\tau$ candidate event}
\label{topology}
Preserving high selection efficiency for the QE $\tau\rightarrow e$ channel at the cost of a larger contamination by external events does not increase in any significant way the size of the sample of interaction vertices located in the target (essentially consisting of  $\nu_\mu$CC and NC events). This number can therefore be predicted by relying only on the size of the essentially uncontaminated 2008 sample. From the 1698 events constituting the 2008 sample,  a fraction of about 5\% of the events was rejected because occurring in bricks equipped with poor quality emulsion films, 1000 interaction vertices have been located in bricks and 110 in dead material. This corresponds to vertex location efficiencies of respectively $74\pm2\%$ and $48\pm4\%$ for CC and NC $\nu_\mu$ interactions. Rescaling for the number of p.o.t., the expected number of located events to be searched for decays of short lived particles in the entire 2008 and 2009 samples is $2978\pm75$. More information on detection efficiencies for signal events is given in Section 5. 

The results presented in this paper come from the decay search analysis of a sample of 2738 events i.e. 92\% of the total statistics. 81\% of the events have an identified muon in the final state. The remaining 8\% of events are currently being processed and the bias this may induce, if any, is negligible

In order to analyse the primary vertex a volume scan is performed over a 1 cm$^2$ area in at least 2 films upstream and 6 films downstream of the vertex lead plate. A procedure has been developed to detect charged and neutral decays as well as secondary interaction and $\gamma$-ray conversion vertices in the vicinity of the primary vertex; it has been introduced in \cite{operacandidate} and detailed information can be found in \cite{decaysearch}.

When a secondary vertex is found, the kinematical analysis of the whole event is performed. This analysis uses the values of the angles measured in the emulsion films, of the momenta determined by multiple Coulomb scattering measured in the brick, of the momenta measured by the magnetic spectrometers, and of the total energy deposited in the instrumented target acting as a calorimeter \cite{operacandidate,eledet,MCS}. The energy of $\gamma$-rays and electrons is estimated by a Neural Network algorithm that uses the combination of the number of track segments in the emulsion films and the shape of the electromagnetic shower, together with the multiple Coulomb scattering of the leading tracks.

This analysis procedure has led to the observation so far of one $\nu_\tau$ candidate event described in detail in \cite{operacandidate}. Among the 7 prongs at the primary vertex, 4 are identified as having been left by a hadron and 3 have a probability lower than 0.1\% of being left by a muon, none being left by an electron. The parent track exhibits a kink topology and the daughter track is identified as left by a hadron through its interaction. Its impact parameter with respect to the primary vertex is ($55\pm4$) $\mu$m where it is smaller than 7 $\mu$m for the other tracks. Two $\gamma$-rays point to the secondary vertex. The event passes all the selection cuts defined in the experiment proposal and summarized in Table \ref{tab:cand} \cite{proposal1}.

\begin{table}[!h]
 \begin{center}
\begin{tabular}{|c|c|c|}
\cline{1-3} Variable &  Cut-off & Value \\
\cline{1-3} Missing P$_T$ at primary vertex (GeV/c) & $<$1.0& $0.57^{+0.32}_{-0.17}$\\
\cline{1-3} Angle between parent track and primary& $>\pi/2$&3.01$\pm$0.03\\
                  hadronic shower in the  & & \\
                   transverse plane (rad)  & & \\
\cline{1-3} Kink angle (mrad) & $>$20 &41$\pm$2\\
\cline{1-3}  Daughter momentum (GeV/c) & $>$2 & $12^{+6}_{-3}$\\
\cline{1-3}  Daughter P$_T$ when $\gamma$-ray  & $>$0.3 & $0.47^{+0.24}_{-0.12}$\\
at the decay vertex (GeV/c)  & & \\
\cline{1-3} Decay length ($\mu$m) & $<$2 lead plates &1335$\pm$35\\
\cline{1-3}
\end{tabular}
\caption{\small{Selection criteria for $\nu_\tau$ candidate events \cite{proposal1}  and values measured for the first observed event \cite{operacandidate}.}} \label{tab:cand}
\end{center}
\end{table}

The invariant mass of the two $\gamma$-rays is ($120\pm20(stat.)\pm35(syst.)$) MeV/c$^2$, consistent with the $\pi^0$ mass. Together with the secondary hadron assumed to be a $\pi^-$ they have an invariant mass of ($640^{+125}_{-80}(stat.)^{+100}_{-90}(syst.)$) MeV/c$^2$. The decay mode is therefore compatible with being $\tau\rightarrow\rho^-(770)\nu_\tau$, the branching ratio of which is about 25\%. The statistical significance of this observation was estimated to be 98.2\%, as the probability that the event was not due to a fluctuation of the background simulating a decay in the $\tau\rightarrow h$ channel \cite{operacandidate}.

 \section{Signal detection efficiencies and physics background}
 \label{kinematics}
\begin{figure}
\begin{center}
  \includegraphics[width=10cm]{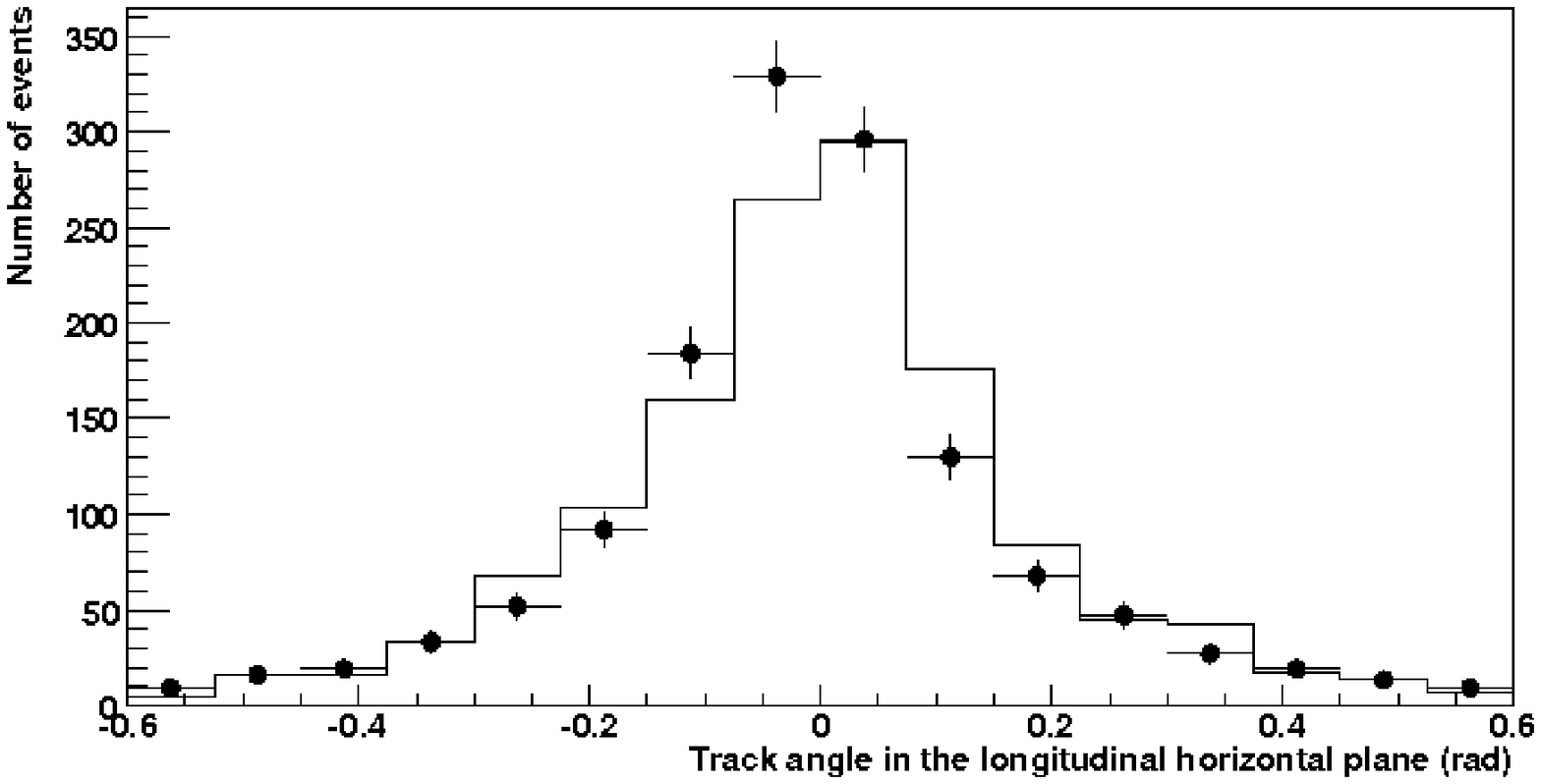}
    \includegraphics[width=10cm]{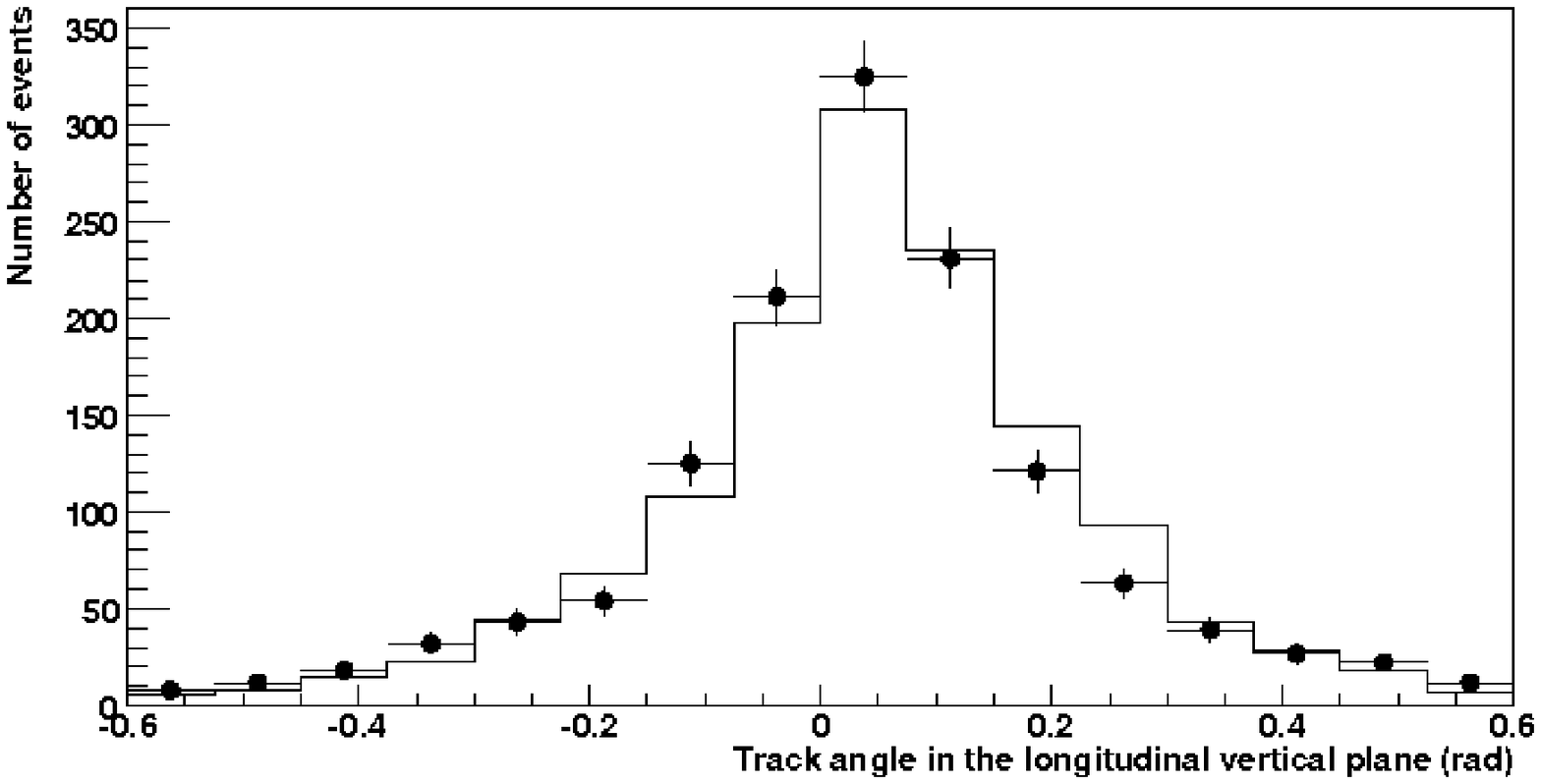}
   \caption{Angle of tracks reconstructed at the primary vertex in the longitudinal horizontal (top) and vertical (bottom) planes: comparison between experimental data (lines with error bars) and simulated data (continuous line).} \label{slopes}
\end{center}
\end{figure}

\begin{figure}
\begin{center}
  \includegraphics[width=10cm]{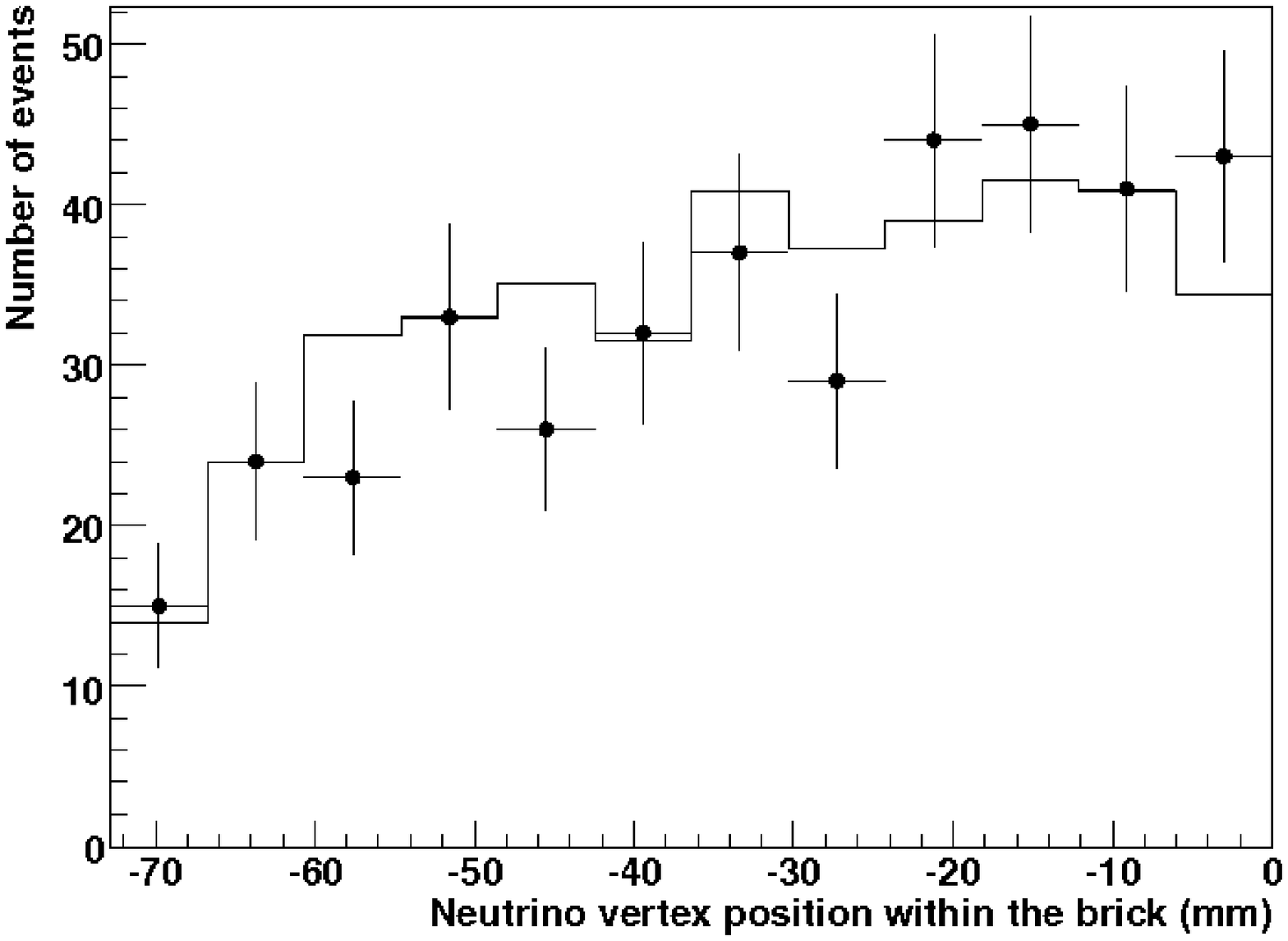}
   \caption{Location of the primary vertex of NC events within the bricks: comparison between experimental data (lines with error bars) and simulated data (continuous line). Most $\nu_\tau$ CC events have topologies similar to NC events.} \label{loca}
\end{center}
\end{figure}

The validity of the Monte Carlo simulation of the electronic detectors response has been verified through a detailed comparison with experimental data. This comparison concerns the muon identification and the reconstruction of its momentum and charge, as well as the reconstruction of the total energy and the hadronic shower profile \cite{eledet}. The event simulation and the off-line reconstruction software have been extended to the response of the emulsion films, allowing all the algorithms used in the analysis of real events to be applied to simulated data. This includes scanning of the CS films, track connection between CS films and downstream films of the brick and track following towards the vertex. The subsequent scan of a volume around the vertex is complemented by track following from the vertex and searching for secondary decay or interaction vertices and for electromagnetic showers \cite{operasimul}. Simulation and off-line reconstruction algorithms have been successfully implemented up to the events topology and their location inside the bricks. The agreement between experimental and simulated data is shown in Figures \ref{slopes} and \ref{loca}. Further work is in progress with the aim of a full, data-driven simulation of all scanning and analysis phases.

\begin{figure}
\begin{center}
  \includegraphics[width=10cm]{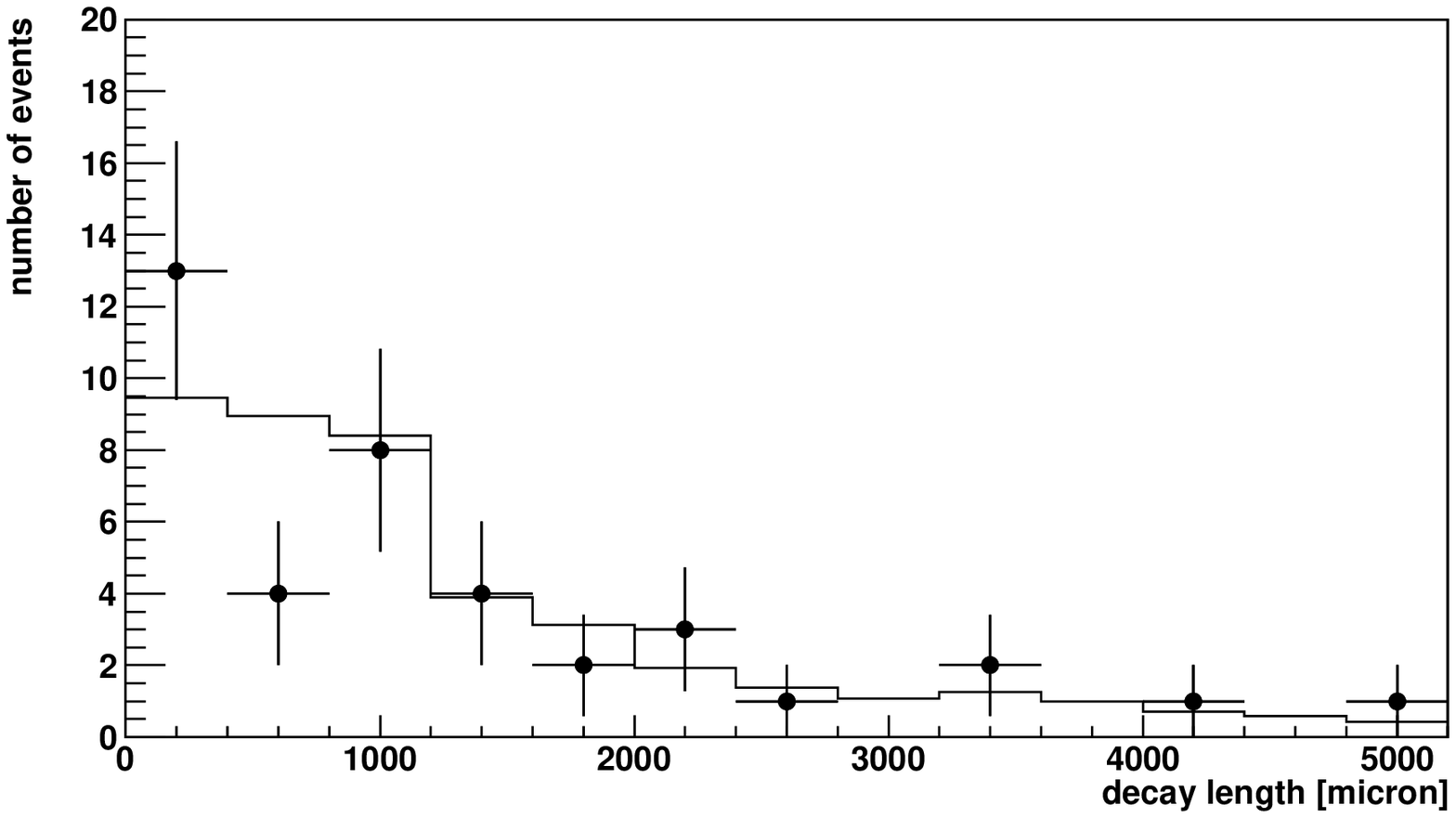}
    \includegraphics[width=10cm]{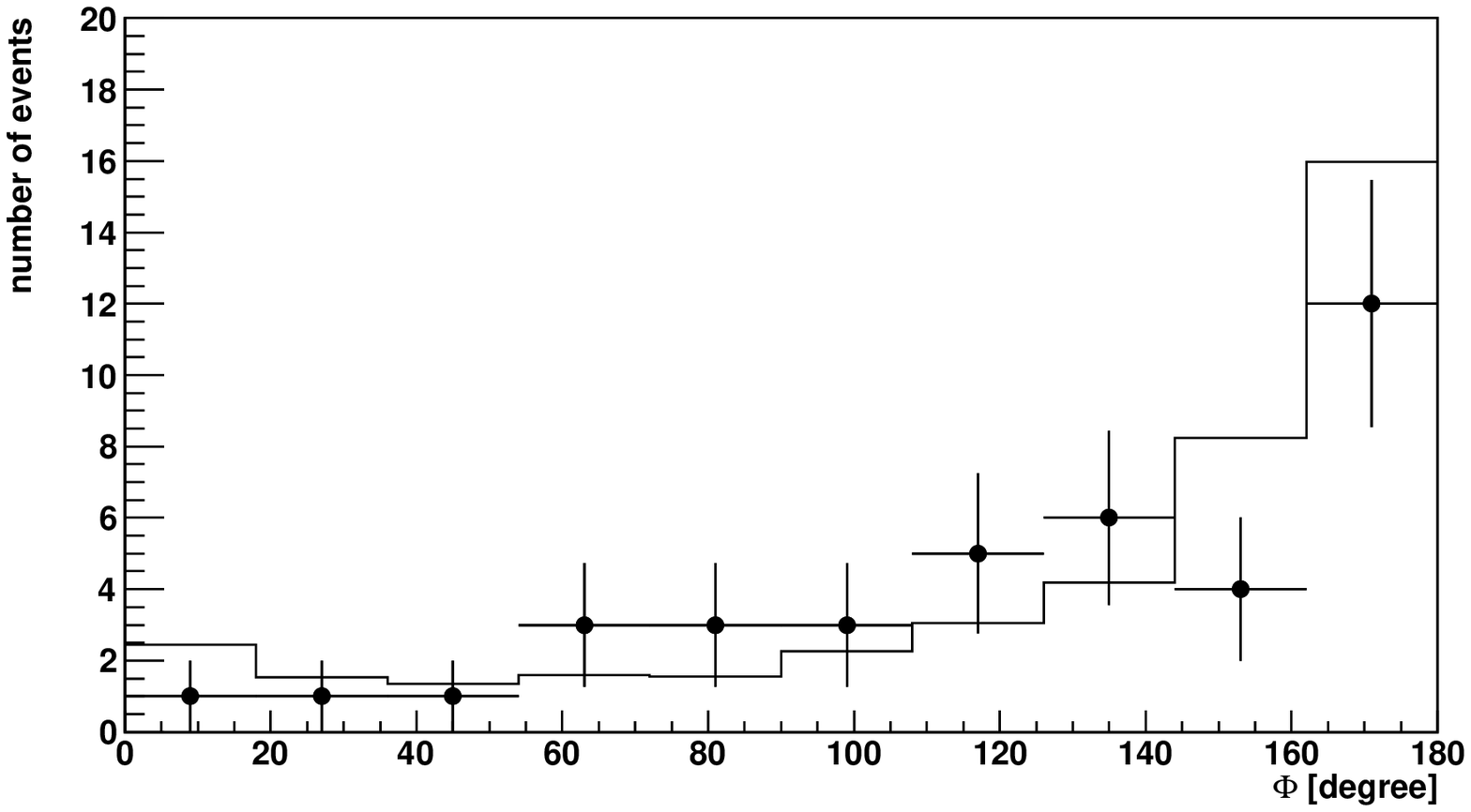}
   \caption{Top: Distributions of the decay length of charmed particles for experimental data (dots with error bars) and simulated data (histogram). Bottom: Distributions of the angle $\phi$ between the parent track and the primary muon in the transverse plane of charmed particles for experimental data (dots with error bars) and simulated data (histogram).} \label{charm}
\end{center}
\end{figure}

Charged charmed particles have lifetimes similar to that of the $\tau$ lepton and share analogous decay topologies. The finding efficiency of the decay vertices is therefore also similar for both types of particles. Comparing the observed charm event sample in size, decay topologies and kinematics with expectations from simulations is thus a straightforward way to verify that prompt-decay selection criteria and their corresponding efficiencies and backgrounds are well understood. Recently published cross-sections have been used in the simulation \cite{operasimul}. The results of this comparison are shown in Table \ref{tab:charm}. Figure \ref{charm}  shows the distributions of the decay length and of the angle $\phi$ between the parent track and the primary muon in the plane transverse to the beam direction. There is a good agreement between experimental and simulated data both in the number of expected charm events and in the above distributions.

\begin{table}[!h]
 \begin{center}
\begin{tabular}{|c|c|c|c|c|}
\cline{1-5} Topology & Observed events & \multicolumn{3}{|c|}{Expected events }\\
& & Charm & Background & Total \\
\cline{1-5} Charged 1-prong & 13 & 15.9 & 1.9 & 17.8 \\
\cline{1-5} Neutral 2-prong & 18 & 15.7 & 0.8 & 16.5 \\
\cline{1-5} Charged 3-prong & 5 & 5.5 & 0.3 & 5.8 \\
\cline{1-5} Neutral 4-prong & 3 & 2.0 & $<$0.1 & 2.1 \\
\cline{1-5} Total & 39 & $39.1\pm7.5$ & $3.0\pm0.9$ & $42.2\pm8.3$ \\
\cline{1-5}
\end{tabular}
\caption{\small{Comparison between charm event topologies observed and expected from simulations including background.}}
\label{tab:charm}
\end{center}
\end{table}

The expected numbers of events in the various $\tau$ channels for the nominal number of $22.5\times 10^{19}$ p.o.t. and for the fraction of the 2008 and 2009 runs analysed so far are shown in Table \ref{tab:tauevent}. Full mixing and $\Delta m^2_{23}=2.5\times10^{-3}$ eV$^2$ are assumed. The total number of signal events expected to be eventually detected has decreased from about 10 as quoted in the experiment proposal \cite{proposal1} to 8, essentially caused by a reduction in the interaction vertex location efficiencies following our more reliable knowledge of the detector and of the analysis procedures. These updated location efficiencies are also shown in Table \ref{tab:tauevent}.

\begin{table}[!h]
 \begin{center}
\begin{tabular}{|c|c|c|c|}
\cline{1-4} Decay channel &  \multicolumn{2}{|c|}{Number of signal events expected for} & Interaction vertex \\
& $22.5\times 10^{19}$ p.o.t. & Analysed sample & location efficiency \\
\cline{1-4} $\tau\rightarrow\mu$ & 1.79 & 0.39 & 0.54 \\
\cline{1-4} $\tau\rightarrow e$ & 2.89 & 0.63 & 0.59 \\
\cline{1-4} $\tau\rightarrow h$ & 2.25 & 0.49 & 0.59 \\
\cline{1-4} $\tau\rightarrow 3h$ & 0.71 & 0.15 & 0.64 \\
\cline{1-4} Total & 7.63 & 1.65 & \\
\cline{1-4}
\end{tabular}
\caption{\small{Expected numbers of observed signal events for $22.5\times 10^{19}$ p.o.t. and for the analysed sample of the data accumulated in the 2008 and 2009 runs. Updated efficiencies for locating interaction vertices appear in the last column}}
\label{tab:tauevent}
\end{center}
\end{table}

The main source of background to all $\tau$ decay channels is constituted by charged charmed particles that decay into similar channels and are produced in $\nu_\mu$CC interactions where the primary muon is not identified. However, the charmed muon decay channel does not contribute to background if the opposite sign muon charge is correctly measured by the spectrometers. Additional charm background comes from $c-\bar{c}$ pair production in NC interactions where one charmed particle is not identified, and from $\bar{\nu}_\mu$, $\nu_e$ and $\bar{\nu}_e$ CC events that amount to 2.5\% in terms of interactions. Second order effects result from the misidentification of the decay products and the topology.

Identifying the muons coming from the primary vertex with the highest possible efficiency is important to suppress background. Details on muon identification algorithms based on signals collected by the electronic detectors can be found in \cite{eledet}; 95\% efficiency is reached for the primary muons of charm events. To further reduce the muon identification inefficiency, all tracks at the primary interaction of signal candidate events emitted with a polar angle $\theta$ smaller than 1 rad are followed within the brick in which the event occurs and from brick to brick. About 30\% of the muons not identified as such by the electronic detectors are recovered through the topology of their end-point, range-momentum correlation and energy loss measurement in the last fraction of their range. The residual inefficiency is dominated by muons emitted at angles larger than 1 rad or escaping the target to the side. The technique in particular allows a sizable reduction of the background in the $\tau\rightarrow\mu$ channel due to wrong associations of the muons emitted in $\nu_\mu$CC events to the vertex of secondary hadronic interactions with kink topologies. This is also true for fake muons in NC events. The hadronic background is also lowered in the $\tau\rightarrow h$ channel due to hadron interactions with kink topologies at the primary vertex of $\nu_\mu$CC events where the primary muon has escaped identification in the electronic detectors. The technique had been applied to the first $\nu_\tau$ candidate event but had however not been taken into account in the background estimate since this has to be assessed via a complex simulation not available at that time. 

The charm background has been evaluated using the charm production cross-sections recently published by the CHORUS Collaboration \cite{chorus}; they are significantly larger than those known at the time of the experiment proposal \cite{proposal1}. The charm production rate induced by neutrinos relative to the charged-current cross-section is 35\% higher at OPERA energies while the relative charm fragmentation fraction into $D^+$ has increased from 10\% to 22\%. The overall effect is a charm background increase by a factor 1.6 to 2.4 depending on the decay channel. 

The second main source of background in the $\tau\rightarrow h$ decay channel comes from one-prong inelastic interactions of primary hadrons produced in NC interactions, or in CC interactions where the primary lepton is not identified and in which no nuclear fragments can be associated with the secondary interaction. This has been evaluated with a FLUKA \cite{FLUKA} based Monte Carlo code, as detailed in \cite{operacandidate} and cross-checked with three sets of measurements. 

\begin{figure}
\begin{center}
  \includegraphics[width=10cm]{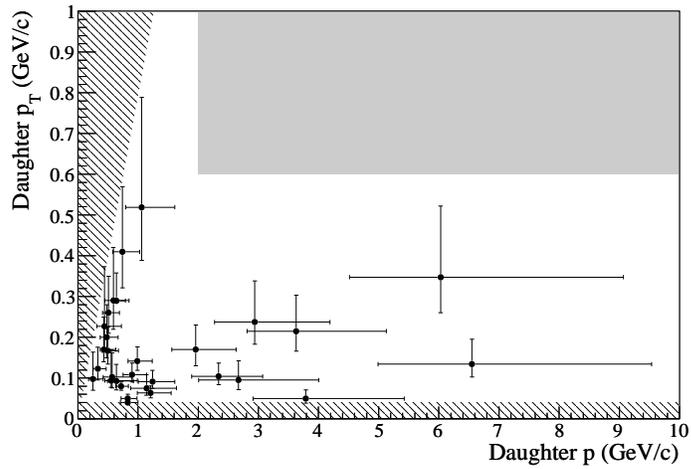}
    \includegraphics[width=10cm]{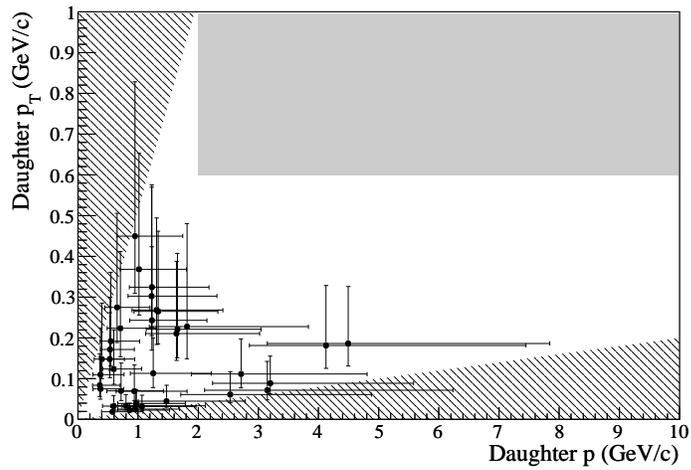}
   \caption{Top: Scatter plot of p$_T$ vs. p of the daughter particle of single prong re-interactions of hadrons far from the neutrino interaction vertices where they are produced. Bottom: Scatter plot of p$_T$ vs. p of the daughter particle of single prong interactions of 4 GeV/c $\pi^-$. On both figures the dark area defines the domain in which $\tau$ decay candidates are selected and the hatched area defines the non-physical region p $<$ p$_T$ and the domain rejected by the selection cuts.} \label{tomoko}
\end{center}
\end{figure}

Tracks of hadrons from neutrino interactions have been followed far from the primary vertex over a total length of 14 m; this corresponds to the total length of tracks left by hadrons in 2300 NC events. No interactions have been found that fulfil the $\tau\rightarrow h$ selection criteria. Immediately outside the signal region, a total of 10 single-prong interactions have been observed with a $p_T$ larger than 200 MeV/c, while 10.8 are expected from simulations. On the top part of Figure \ref{tomoko}  the total momentum is plotted versus the transverse momentum of the final state particles for these interactions. The parameters space in which $\tau$-decay candidates are accepted is shown. 

\begin{figure}
\begin{center}
  \includegraphics[width=6.5cm]{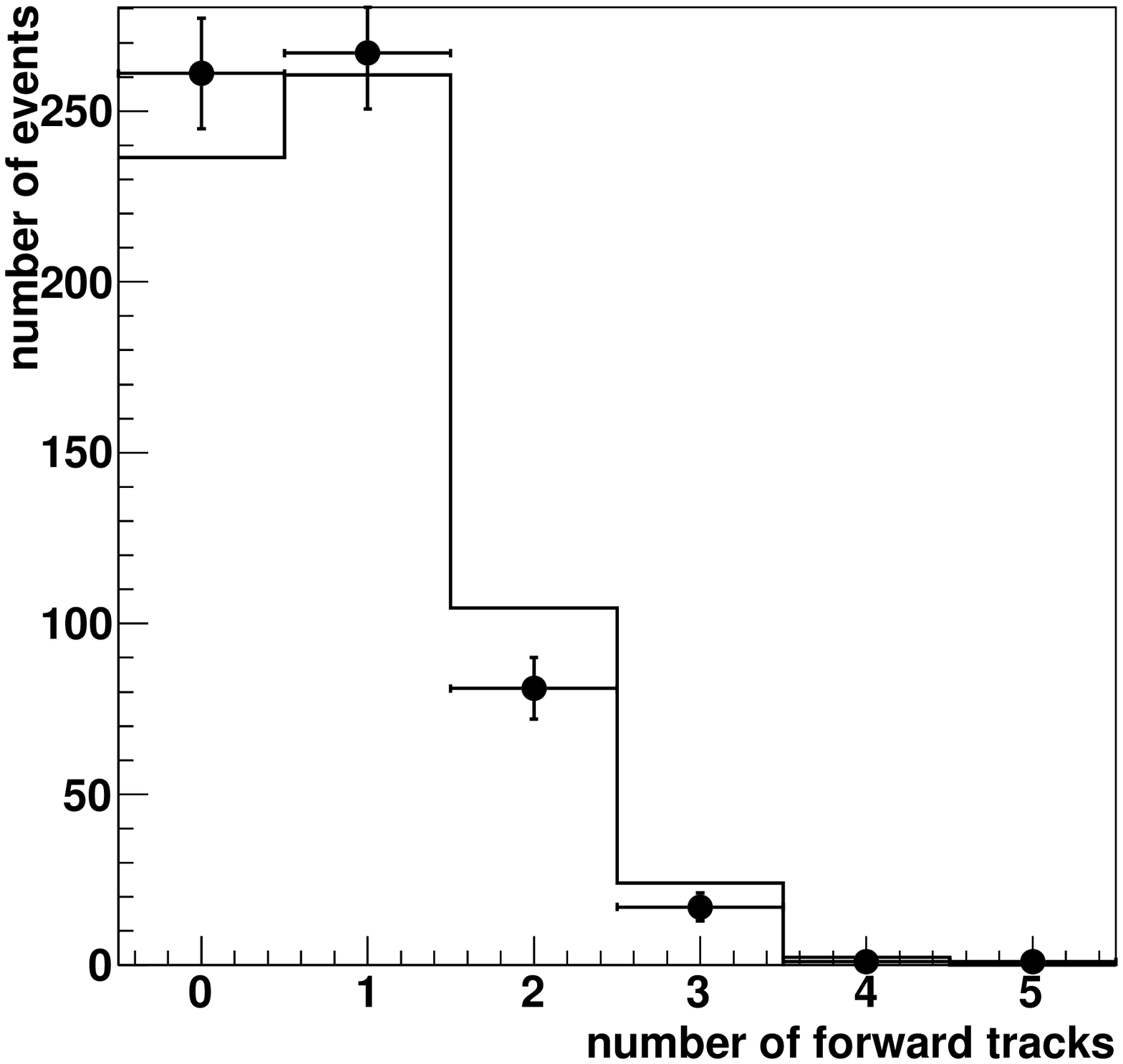}
    \includegraphics[width=6.5cm]{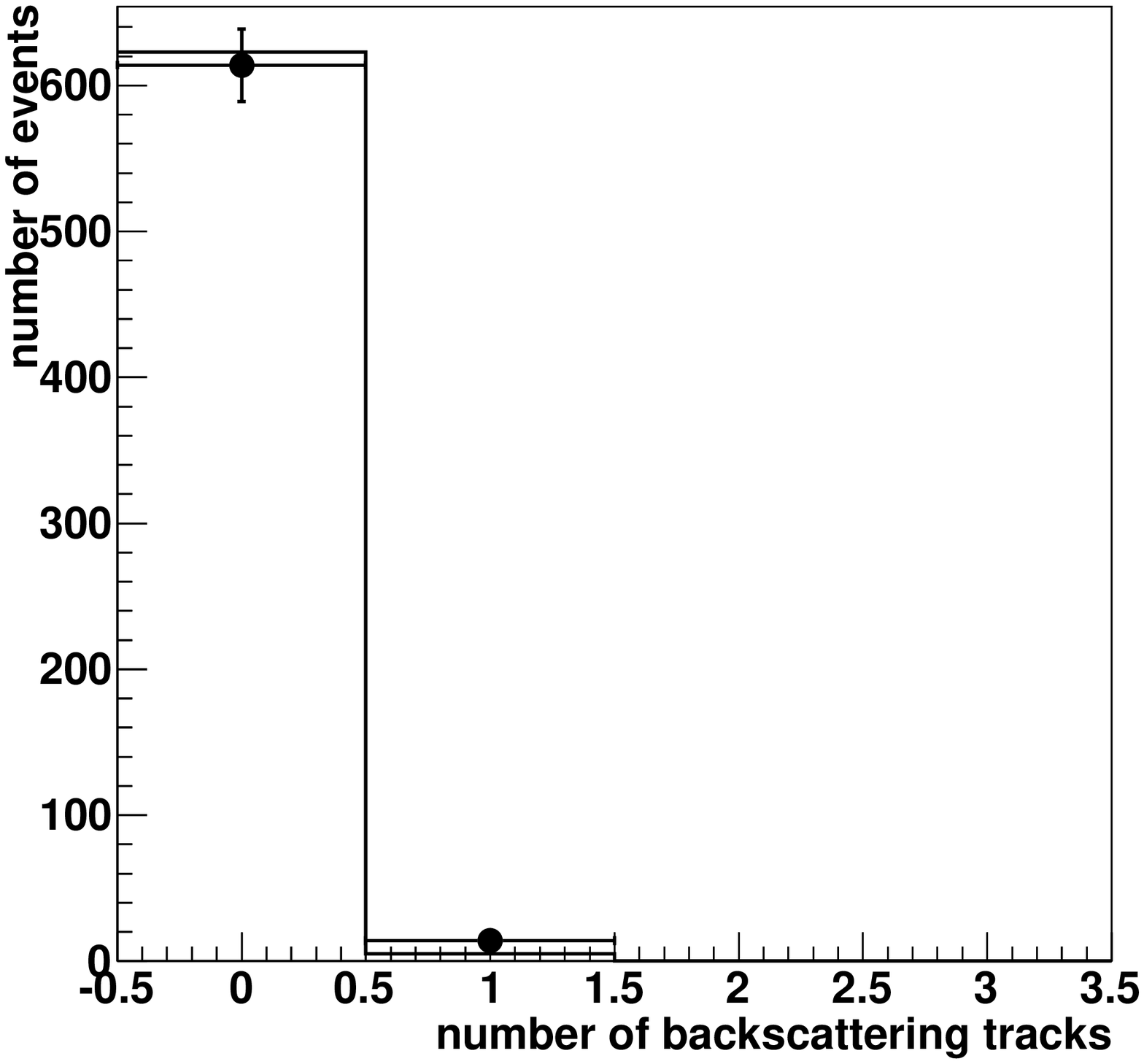}\\
      \includegraphics[width=6.5cm]{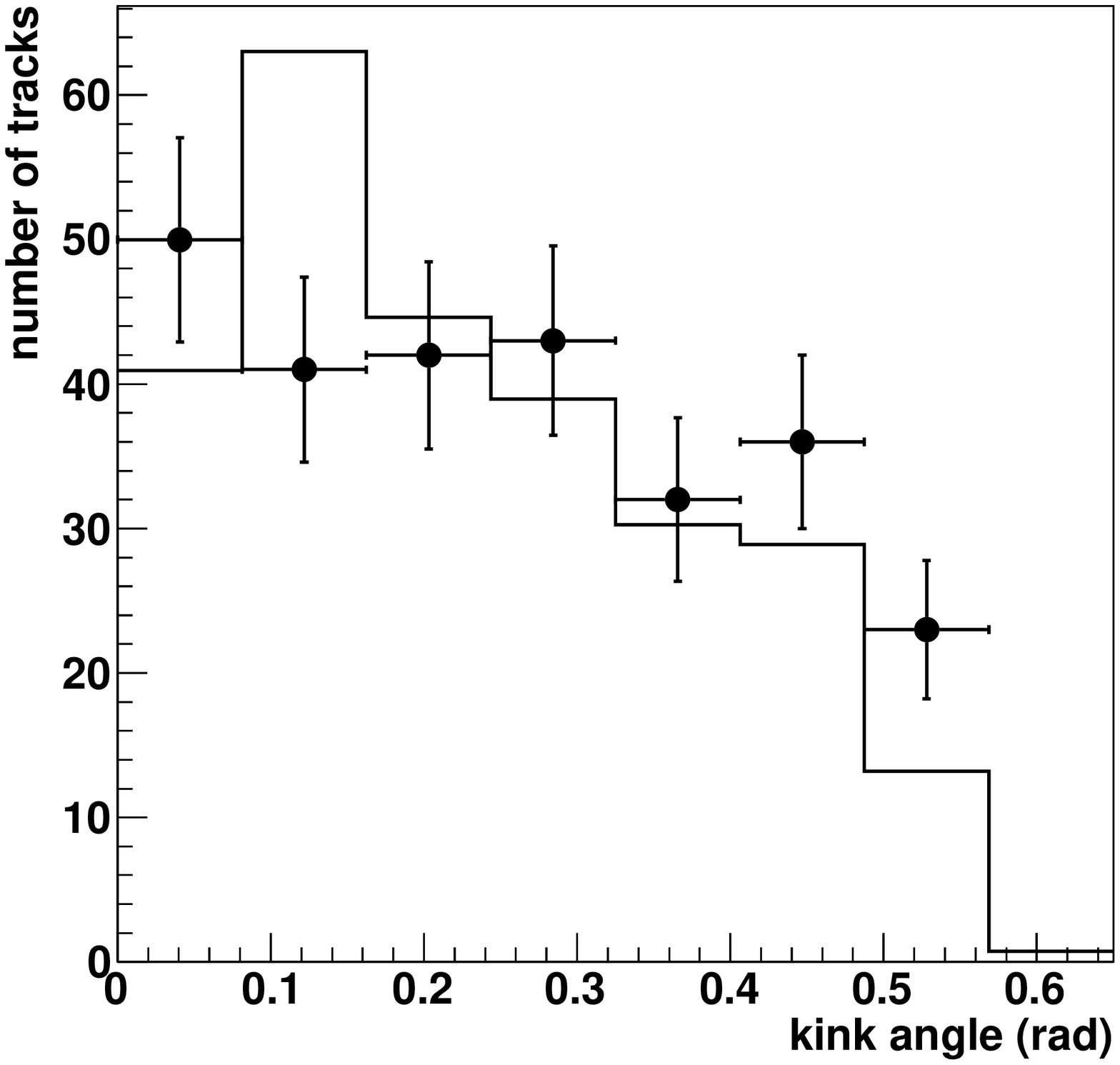}
    \includegraphics[width=6.5cm]{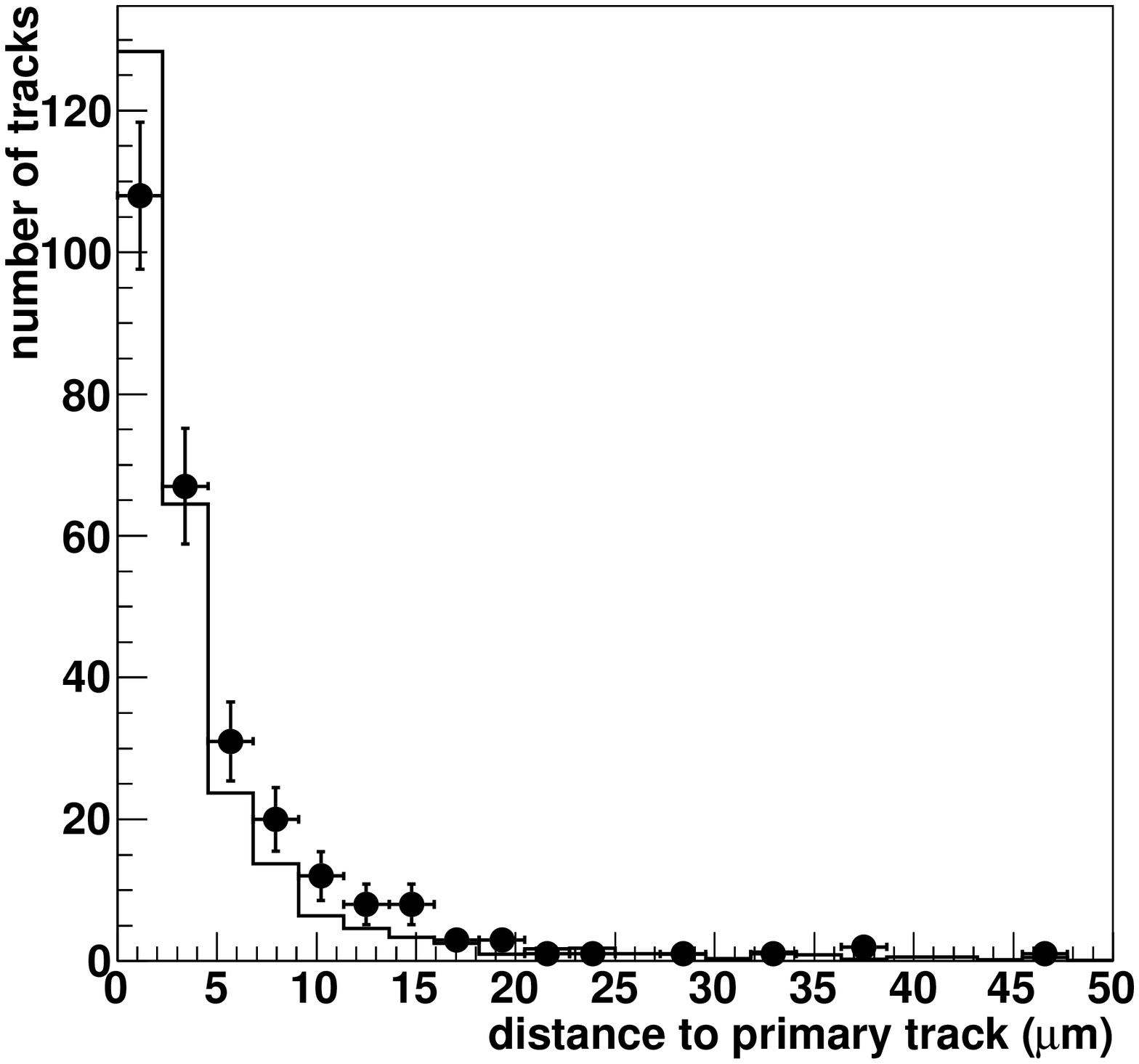}
   \caption{Comparison between experimental data (dots with error bars) and simulation data (histogram) for two bricks exposed to 4 GeV/c $\pi^-$ test beams. Normalisations are independent. Top left: forward tracks multiplicity. Top right: backward tracks multiplicity. Bottom left: kink angle for one-prong events. Bottom right: minimum distance between the primary and the secondary tracks for one-prong events.} \label{pion}
\end{center}
\end{figure}

Bricks of the OPERA type exposed to pion beams have also been studied in order to further crosscheck the estimate of the hadron interaction background. In two test beam exposures with 4 GeV/c $\pi^-$ a total of about 190 m of track length have been followed. In the first exposure, 314 interactions were localised with an angular acceptance $\theta<1$ rad, out of which 126 are single-prong events with a kink angle larger than 20 mrad (the same selection cut as for the $\tau$ decay search). In the second exposure, 220 interactions were located with an angular acceptance $\theta<0.54$ rad, out of which 88 are single-prong events with kink angles larger than 20 mrad. On the bottom part of Figure \ref{tomoko}  the transverse momentum is plotted versus the total momentum of the final state particles for the single-prong events in this second exposure. Comparisons showing a fair agreement between experimental and simulated data of both exposures are presented in Figure \ref{pion} for a set of topological variables. 

\begin{figure}
\begin{center}
  \includegraphics[width=10cm]{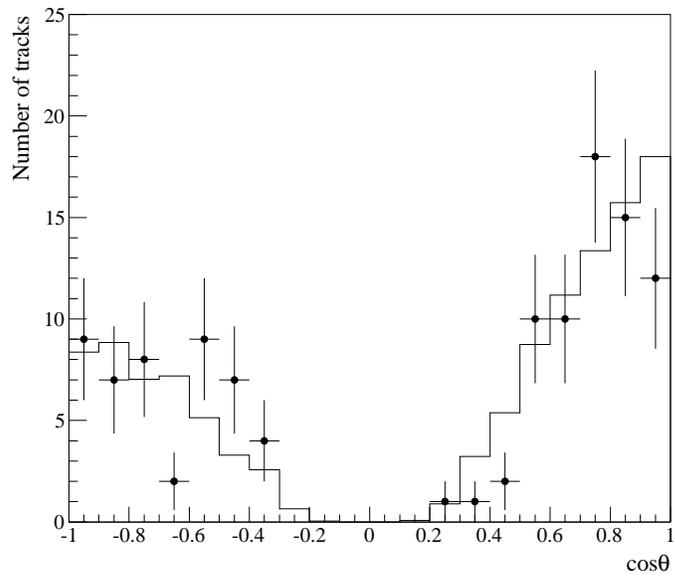}
  \caption{Polar angle distributions of the highly ionising particles emitted in 8 GeV/c $\pi^-$ interactions in an OPERA brick for experimental data (dots with error bars) and simulated data (histogram). Forward tracks correspond to $\cos\theta=1$.} \label{black}
\end{center}
\end{figure}

The hadron interactions background can be further reduced by increasing the detection efficiency of highly ionizing particles, low energy protons and nuclear fragments, emitted in the cascade of intra-nuclear interactions initiated by the primary particles and in the nuclear evaporation process. This is a novel feature not yet implemented in the analysis reported in \cite{operacandidate} . In order to detect a significant fraction of those fragments emitted at large angle, an image analysis tool has been developed where 2.5 mm$\times$ 2.1 mm $\times$ 24 layers of high resolution microscope tomographic images are analysed in the upstream and downstream films of an interaction vertex. This technique has been used to study a sample of 64 interactions in a brick of the OPERA type exposed to an 8 GeV/c $\pi^-$ beam. At least one highly ionising particle has been associated to ($56\pm7$)\% of the events with a backward/forward asymmetry of $0.75\pm0.15$, while 53\% are expected from simulations with an asymmetry of 0.71. Figure \ref{black} shows the fair agreement in polar angle distribution of the highly ionising particles between experimental and simulated data. The technique allows detecting more highly ionising particles associated to secondary vertices. It provides an additional background reduction of about 20\%. No such particles have been found to be associated to the decay vertex of the first $\nu_\tau$ candidate event.

The expected background in the muon decay channel caused by large angle muon scattering has been evaluated in \cite{proposal1}.

The total number of expected background events has slightly decreased from 0.75 as quoted in the experiment proposal \cite{proposal1} to 0.73, despite a significant increase of the charm cross-sections, essentially thanks to a significant improvement in the identification of the decay products as hadron or muon. All background sources are summarized in Table \ref{tab:total}. Systematic errors of 25\% on charm background and of 50\% on hadron and muon backgrounds are assumed. Errors arising from the same source are combined linearly and otherwise in quadrature.

\begin{table}[!h]
\begin{center}
\begin{tabular}{|c|c|c|c|c|c|c|c|c|}
\cline{1-9} Decay &  \multicolumn{8}{|c|}{Number of background events expected for} \\
channel & \multicolumn{4}{|c|}{$22.5\times 10^{19}$ p.o.t. } & \multicolumn{4}{|c|}{Analysed sample }  \\
& Charm & Hadron & Muon & Total & Charm & Hadron & Muon & Total \\
\cline{1-9} $\tau\rightarrow\mu$ & 0.025 & 0.00 & 0.07 & $0.09\pm0.04$ & 0.00 & 0.00 & 0.02 & $0.02\pm0.01$ \\
\cline{1-9} $\tau\rightarrow e$ & 0.22 & 0.00 & 0.00 & $0.22\pm0.05$ & 0.05 & 0.00 & 0.00 & $0.05\pm0.01$ \\
\cline{1-9} $\tau\rightarrow h$ & 0.14 & 0.11 & 0.00 & $0.24\pm0.06$ & 0.03 & 0.02& 0.00 & $0.05\pm0.01$ \\
\cline{1-9} $\tau\rightarrow 3h$ & 0.18 & 0.00 & 0.00 & $0.18\pm0.04$ & 0.04 & 0.00& 0.00 & $0.04\pm0.01$ \\
\cline{1-9} Total & 0.55 & 0.11 & 0.07 & $0.73\pm0.15$ & 0.12 & 0.02& 0.02 & $0.16\pm0.03$ \\
\cline{1-9}
\end{tabular}

\caption{\small{Expected numbers of observed background events from different sources for the nominal number of $22.5\times 10^{19}$ p.o.t. and for the analysed sample of the data accumulated in the 2008 and 2009 runs. Errors quoted are systematic.}}
\label{tab:total}
\end{center}
\end{table}

\section{Signal statistical significance}
\label{background}
One $\nu_\tau$ candidate event is observed in the $\tau\rightarrow h$ decay channel that passes all the selection cuts where $0.49\pm12$ events are expected for this decay mode in the currently analysed sample assuming full mixing and $\Delta m^2_{23}=2.5\times10^{-3}$ eV$^2$. The background in this channel is estimated to $0.05\pm0.01(syst.)$ event. The probability for the event not to be due to a background fluctuations and thus the statistical significance of the observation is 95\%.  Considering all decay channels, the numbers of expected signal and background events are respectively $1.65\pm0.41$ and $0.16\pm0.03(syst.)$, the probability for the event to be background being 15\%.

\section{Conclusions}
\label{conclu}
The OPERA experiment has completed the study of 92\% of the data accumulated during the first two years of operation in the CNGS beam (2008-2009) with the aim of the first detection of neutrino oscillations in direct appearance mode. 

The observation so far of a single candidate $\nu_\tau$ event is compatible with the expectation of 1.65 signal events. The significance of the observation of one decay in the $\tau\rightarrow h$ channel has decreased from 98.2\%  in the first analysis \cite{operacandidate} to 95\%, as the size of the analysed event sample has substantially augmented.

The increase in the anticipated background resulting from the larger charm cross-sections recently measured \cite{chorus} compared to those known at the time of the experiment proposal has been more than compensated by a higher muon identification efficiency. In addition, the study of highly ionising tracks left by protons and nuclear fragments which are often associated to hadronic re-interactions has allowed reducing by about 20\% this background specific to the hadronic decay modes of the $\tau$. 

The event location efficiency has been re-evaluated for several phases of the analysis procedure. Completing the extension of the events simulation and their reconstruction in the emulsion films of the target bricks, a task already well advanced down to the vertex reconstruction, will help in further controlling the detection efficiencies at each step of the analysis process, from the track reconstruction in the electronic detectors to the location of primary and secondary vertices in the emulsion films. This will also allow reevaluating the selection criteria in view of improving the signal-to-noise ratio. 

The analysis of the large event samples collected in the 2010 and 2011 CNGS runs and corresponding to $6.90\times10^{19}$ p.o.t. at the moment of writing is in progress.  

\section{Acknowledgements}
We thank CERN for the successful operation of the CNGS facility and INFN for the continuous support given to the experiment during the construction, installation and commissioning phases through its LNGS laboratory. We warmly acknowledge funding from our national agencies: Fonds de la Recherche Scientifique - FNRS and Institut Interuniversitaire des Sciences Nucl{\'e}aires for Belgium, MoSES for Croatia, CNRS and IN2P3 for France, BMBF for Germany, INFN for Italy, JSPS (Japan Society for the Promotion of Science), MEXT (Ministry of Education, Culture, Sports, Science and Technology), QFPU (Global COE program of Nagoya University, ÓQuest for Fundamental Principles in the UniverseÓ supported by JSPS and MEXT) and Promotion and Mutual Aid Corporation for Private Schools of Japan, SNF, the University of Bern and ETH Zurich for Switzerland, the Russian Foundation for Basic Research (grant 09-02-00300\_a), the Programs of the Presidium of the Russian Academy of Sciences ÓNeutrino PhysicsÓ and ÓExperimental and theoretical researches of fundamental interactions connected with work on the accelerator of CERNÓ, the Programs of support of leading schools (grant 3517.2010.2), and the Ministry of Education and Science  of  the Russian Federation for Russia., the Korea Research Foundation Grant (KRF-2008-313-C00201) for Korea and TUBITAK, The Scientific and Technological Research Council of Turkey, for Turkey. We are also indebted to INFN for providing fellowships and grants to non-Italian researchers. We thank the IN2P3 Computing Centre (CC-IN2P3) for providing computing resources for the analysis and hosting the central database for the OPERA experiment. We are indebted to our technical collaborators for the excellent quality of their work over many years of design, prototyping and construction of the detector and of its facilities.








\end{document}